\documentclass[reprint, amsmath,amssymb, twocolumn]{revtex4}
\usepackage{graphicx}
\usepackage{amsmath}

\begin{document}
\title{Broadband parametric amplifiers based on nonlinear kinetic inductance artificial transmission lines}

\author {S. Chaudhuri$^1$}
\author {D. Li$^{2}$}
\author {K. D. Irwin$^{1,2}$}
\author{C. Bockstiegel$^3$}
\author{J. Hubmayr$^4$}
\author{J. N. Ullom$^4$}
\author{M. R. Vissers$^4$}
\author{J. Gao$^4$}

\affiliation{
1) Department of Physics, Stanford University, Stanford, CA 94305, USA\\
2) SLAC National Accelerator Laboratory, Menlo Park, CA 94025, USA\\
3) Department of Physics, University of California, Santa Barbara, CA 93106, USA\\
4) National Institute of Standards and Technology, Boulder, CO 80305, USA}

\date{\today}

\begin{abstract}

We present broadband parametric amplifiers based on the kinetic inductance of superconducting NbTiN thin films in an artificial (lumped-element) transmission line architecture. We demonstrate two amplifier designs implementing different phase matching techniques: periodic impedance loading, and resonator phase shifters placed periodically along the transmission line. Our design offers several advantages over previous CPW-based amplifiers, including intrinsic 50 ohm characteristic impedance, natural suppression of higher pump harmonics, lower required pump power, and shorter total trace length. Experimental realizations of both versions of the amplifiers are demonstrated. With a transmission line length of 20 cm, we have achieved gains of 15 dB over several GHz of bandwidth.
\end{abstract}

\maketitle

Cryogenic low-noise broadband amplifiers are critical for a variety of applications, such as the multiplexed readout of astronomical detectors \cite{zmuidzinas_2012, zmuidzinas_2003, irwin_2004} and superconducting qubits \cite{schoelkopf_2004, martinis_2012, schoelkopf_2013}, the manipulation of mechanical resonators coupled to microwave cavities \cite{lehnert_2009}, and the study of nonclassical states of microwave light \cite{lehnert_2008, wallraff_2011}. These experiments often use high electron mobility transistor (HEMT) amplifiers, which typically have a noise temperature of 2-5 K in the 4-8 GHz range\cite{lnf}. This noise is 10-40 times above the standard quantum limit,  the fundamental limit imposed by quantum mechanics\cite{caves}.

Over the last decade, there has been rapid development of quantum-limited microwave amplifiers, including particularly Josephson parametric amplifiers (JPAs) \cite{lehnert_2007, devoret_2010, siddiqi_2011}. JPAs use the dissipationless nonlinearity of Josephson junctions in a parametric process to achieve gain. Both narrowband JPAs, based on junction-embedded resonant architectures, and broadband JPAs, based on junction-embedded transmission lines, have been developed and have demonstrated near quantum-limited noise performance\cite{siddiqi_2014, siddiqi_2015}. However, the $\sim$ 10 $\mu$A critical current of the junctions limits the dynamic range of these devices and excludes them from some important applications, such as the multiplexed readout of thousands of qubits or detectors. In addition, fabricating a large number ($>1000$) of junctions with high yield is a nontrivial task. 

Recently, another type of broadband parametric amplifier based on the nonlinear kinetic inductance of NbTiN transmission lines has been proposed \cite{zmuidzinas_2012b, pappas_2013, chaudhuri_2014,pappas_2016, danilov_2016}. These devices are simple to fabricate, requiring only one lithography step (patterning of the NbTiN film). Due to the $\gtrsim$1 mA critical currents of the films, the amplifier saturation power is 5-6 orders of magnitude higher than JPAs, making them promising for readout of a large array of detectors or qubits. In contrast with reflection-type resonant JPAs, the traveling-wave architecture of the NbTiN amplifier eliminates the need for a circulator, enabling on-chip integration with a detector or qubit. In previous work, NbTiN amplifiers were realized as long coplanar waveguide (CPW) transmission lines, with $>$20 dB gain over several GHz bandwidth\cite{pappas_2013}. Despite the excellent gain performance, the CPW amplifier has a few drawbacks. To achieve $\gtrsim$15 dB gain, a CPW over 2 meters long is used, which can result in low fabrication yield. The characteristic impedance $Z_0$ is 200 ohms, requiring an impedance transformer to match to 50 ohm external circuitry; impedance mismatch is the likely cause of large ripples in the gain profile. Associated with the large $Z_0$, the pump power, determined by $P_{p}=I_{rms}^{2}Z_{0}$ where $I_{rms} \sim$ 1 mA is the rms pump current, 
is $\sim$200 $\mu$W. As a result, the amplifier chip heats up, likely creating thermal noise. Consequently, quantum-limited performance has yet to be demonstrated in these devices. 

In this letter, we present an alternative amplifier architecture that may solve these problems: an artificial transmission line. We present the design and experimental data from two amplifier architectures, and in both versions, we demonstrate $\sim$15 dB gain over a few GHz of bandwidth.

As demonstrated in previous work\cite{zmuidzinas_2012b,pappas_2013,gao_2016}, NbTiN films have nonlinear, current-dependent kinetic inductance with extremely low dissipation, $L(I)=L_{0}(1+ I^2/I_{*}^2)$, equivalent to an optical Kerr media. $I_{*}$, the scale of the nonlinearity, is on the order of $\sim$10 mA, resulting in amplifiers with high saturation power. The inductance nonlinearity gives rise to four-wave mixing in the parametric amplifier. Two pump photons from a strong tone are converted into a signal photon and an idler photon, producing gain. There are three aspects critical to a NbTiN traveling-wave parametric amplifier design: the use of nonlinear transmission line as the gain media, phase-matching structures to achieve high (exponential) gain over a broad bandwidth, and structures to suppress higher harmonics of the  pump\cite{landauer_1963,day_2014,siddiqi_2014,chaudhuri_2014}. 

In our design, an artificial transmission line made of NbTiN serves as the gain medium. An artificial transmission line, also known as a lumped-element transmission line, uses lumped-element inductors and capacitors instead of the distributed inductances and capacitances in a conventional transmission line.\cite{pozar} A lumped-element transmission line with inductance $L$ and capacitance $C$ per unit cell possesses a characteristic impedance $Z_{0}=\sqrt{L/C}$ and a low-pass cutoff frequency $f_{c}=1/\pi\sqrt{LC}$. At frequencies far below $f_c$, an artificial transmission line behaves like a conventional transmission line with the same characteristic impedance $Z_0$. A 50 $\Omega$ artificial transmission line may readily be made by appropriately tuning the unit-cell inductance and capacitance, eliminating the need for impedance transformers used in the CPW design. The intrinsic low-pass characteristic also naturally suppresses any pump harmonics above $f_c$. 

We present two amplifier designs implementing different phase-matching architectures: the periodic impedance loading scheme, and the resonator phase shifter scheme. The two designs also use different harmonic suppression methods.

\begin{figure}[ht]
\includegraphics[width=8.5cm]{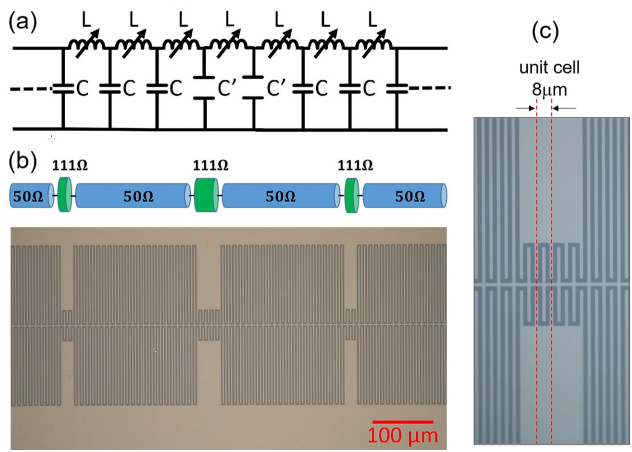}
\caption{Design of the periodically-loaded NbTiN ``fishbone" parametric amplifier. (a) Schematic circuit model for a periodic loading section in an artificial transmission line, in which the capacitance is reduced from C to C'. 
(b) Micrograph of periodically loaded artificial transmission line showing one period of the loading section (bottom) and a cartoon showing its transmission line model (top). (c) A zoomed-in view highlighting an LC unit cell in the loaded section. In both (b) and (c), light color represents areas covered by NbTiN and dark color represents gap areas with NbTiN etched off.}
\label{fig:fishbone_design}
\end{figure}

The design with periodic impedance loadings is shown in Fig.~\ref{fig:fishbone_design}. We place 2 $\mu$m interdigitated capacitor (IDC) fingers on both sides of the 2$\mu$m center strip, forming a ``fishbone" pattern as shown in Fig. \ref{fig:fishbone_design}(b,c). The inductance and capacitance per unit cell (8~$\mu$m long) are 50 pH and 20 fF, respectively, yielding a characteristic impedance of 50 ohms and a cutoff frequency of $f_{c}=318$ GHz. 

The loading scheme in Fig. \ref{fig:fishbone_design}(a,b) is analogous to that used in CPW-based NbTiN amplifiers\cite{zmuidzinas_2012b, pappas_2013}. The characteristic impedance of the transmission line is modified every one-sixth of a wavelength at a frequency slightly above the pump frequency $f_{p}$ to form a wide stopband at 3$f_p$. In addition, every third loading is modified in length (longer or shorter relative to the first two) to create a narrow stopband near $f_p$. Placed just below this narrow stopband, the pump tone picks up additional phase shift to fulfill the phase-matching condition, while its third harmonic is suppressed by the 3$f_p$ stopband. As a result, exponential gain can be achieved over a broad bandwidth of more than one-half the pump frequency\cite{zmuidzinas_2012b,chaudhuri_2014}.

We create periodic impedance loadings by reducing the length of the IDC fingers in the loaded sections. As shown in Fig. \ref{fig:fishbone_design}(c), every 22 unit cells (corresponding to one-sixth of a wavelength at $\sim$6 GHz), the IDC finger length -- and thus, the capacitance -- is reduced by a factor of 5. This effectively increases the characteristic impedance from 50 ohms to 111 ohms. The reduced capacitance occurs for two consecutive unit cells. In every third loading, the number of unit cells with reduced capacitance is doubled to four. The cutoff frequency of 318 GHz is sufficiently high that it does not interfere with the loading stopbands for the pump and third harmonic. The electrical length of the amplifier is approximately 70 wavelengths at the pump frequency $\sim$ 6 GHz, and the physical length is 10 cm.



\begin{figure}[ht]
\includegraphics[width=8.5cm]{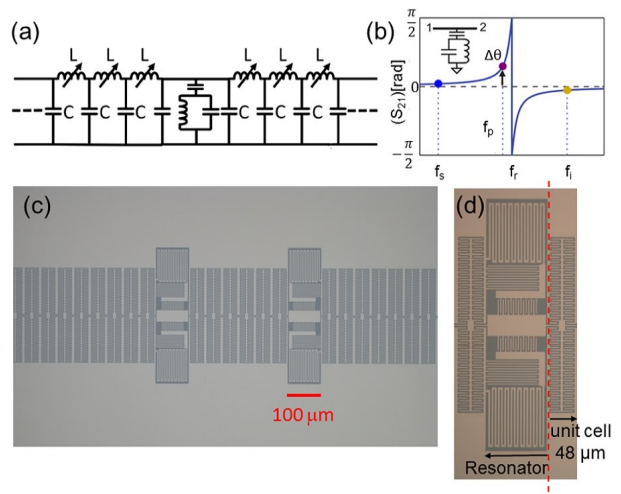}
\caption{Design of the LC resonator-embedded NbTiN ``leaf" parametric amplifier. (a) Schematic circuit model for the resonator embedding in an artificial transmission line. (b) Phase shift through a resonator. The resonator introduces a frequency-dependent phase shift which acts mainly on the pump tone. (c) Micrograph of a section of the device showing a 4-resonator phase shifter block embedded in the transmission line. (d) A zoom-in view highlighting a symmetric resonator pair and a unit cell. In both (c) and (d), light color represents areas covered by NbTiN and dark color represents gap areas with NbTiN etched off.}
\label{fig:leaf_design}
\end{figure}

Fig. \ref{fig:leaf_design}(a) shows an alternative phase-matching scheme based on resonator phase shifters embedded along the transmission line. This technique was demonstrated earlier for traveling-wave Josephson parametric amplifiers in \cite{martinis_2015}. To correct for the nonlinearity-induced phase mismatch, we periodically place resonators, with resonant frequency slightly above the desired pump $f_r \gtrsim f_p$, along the transmission line. The resonator phase shifts, but does not reflect, the pump tone while leaving the phase of the signal and idler far detuned from the resonator unchanged (see Fig. \ref{fig:leaf_design}b). In this manner, the phase mismatch accumulated along each section of nonlinear transmission line is countered by the phase shift introduced by the resonator. For a sufficient number of resonators, the structure approximates a transmission line with continuous dispersion, and phase-matched gain may be achieved\cite{martinis_2015}.

In our design in Fig. \ref{fig:leaf_design}(c,d), which resembles the shape of a leaf, the inductance and capacitance in each unit cell of the lumped-element transmission line (displayed schematically in Fig. \ref{fig:leaf_design}a) are 290 pH and 116 fF, yielding a characteristic impedance of 50 ohms and a cutoff frequency of $f_{c}=55$ GHz. Because the third harmonic of the pump tone at $3f_p \approx$ 18 GHz is a significant fraction of $f_{c}$, simulation shows that the harmonic growth is limited by the natural phase mismatch between the pump and third harmonic due to the intrinsic dispersion from the low-pass cutoff limits. This option represents an advantage of the lumped-element architecture over the CPW. Every 340 unit cells, a set of four resonators is embedded in the transmission line. These resonators have a design resonant frequency of $6$ GHz and a quality factor of 70. Two resonators are embedded in parallel at the same point symmetric about the center strip, as shown in Fig. \ref{fig:leaf_design}c. This symmetry prevents the excitation of parasitic slot-line modes. Following an additional six unit cells, equivalent to one-quarter wavelength at $f_p$, a second pair of resonators is embedded. The four-resonator section serves as a phase shifter block. Simulation shows that each block can introduce up to 30 degree pump phase shift with negligible insertion loss ($< 0.1$~dB).
A total of six resonator phase shifter blocks are embedded in the 10 cm-long transmission line. The electrical and physical lengths are approximately the same as for the ``fishbone" design.  

Both amplifier designs were fabricated on a 6-inch intrinsic silicon wafer. A NbTiN film 20 nm thick was deposited on the Si substrate and patterned using a CF$_4$-O$_2$ etch to minimize overetch into the Si and maintain a constant impedance over the 6-inch wafer. The critical temperature of the film was measured to be 15 K. The wafer was diced into 10 cm long and 4 - 10 mm wide chips with one or a few devices on each chip. 

We cooled the devices to a temperature of 100 mK in a dilution refrigerator and measured the broadband gain using a vector network analyzer (VNA). The gain was determined by measuring the transmission through the device with and without the presence of the pump tone. 




\begin{figure}[htp]
\vspace{-1.75cm}
\includegraphics[width=8.5cm]{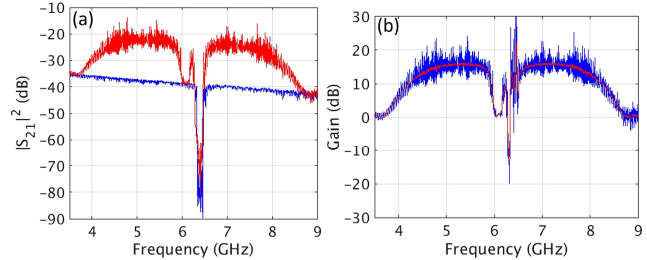}
\vspace{-1.75cm}
\caption{\small Gain-bandwidth measurements from two cascaded ``fishbone" amplifiers with periodic impedance loadings. (a) Transmission $|S_{21}|^{2}$ with the pump off (blue) and the pump on (red). (b) Gain profile. The measured gain curve is shown in blue, while a smoothed gain curve is shown in red.} \label{fig:fishbone_gain}
\end{figure}

We first measured a single 10-cm amplifier device and achieved a maximum gain of 9 dB in both versions. In order to achieve our target gain of 15dB, for both designs, we cascaded two amplifiers which effectively created a 20 cm-long device.

The measurement result from the cascaded ``fishbone" amplifiers is shown in Fig. \ref{fig:fishbone_gain}. The transmission through the device with the pump off (blue) and the pump on (red) is illustrated in Fig. \ref{fig:fishbone_gain}a. 
The stopband formed by the periodic loading structure that introduces the pump dispersion falls at 6.4 GHz, which agrees well with our circuit simulation. We achieve broadband gain (red) by injecting a pump tone at $f_p = 6.22$~GHz with a power of 100 $\mu$W. Fig. \ref{fig:fishbone_gain}b demonstrates a peak gain of 15 dB, with a 3 dB double-sided bandwidth of 3 GHz (1.5GHz on each side of the pump frequency). There are two dips in the gain profile next to the pump frequency. The dip above (below) the pump frequency is due to the signal (idler) falling in the stopband. Comparable gain and bandwidth has only been achieved with CPW amplifiers having an order of magnitude greater length (2 m) and a factor of 2 greater pump power (200 $\mu$W)\cite{zmuidzinas_2012b, pappas_2013}. The gain ripple at peak gain, $\pm3$ dB, is also improved from the previous CPW amplifier. We attribute this to the ``engineered" 50 ohm characteristic impedance of the artificial transmission line and the elimination of impedance transformers. Indeed, time domain reflectometry  (TDR) measurements on the device have verified that $Z_0 = 47\Omega$. TDR measurements also reveal that some reflection remains at the input/output ports, likely from wirebonds. We plan to improve the connection technique in the future.

\begin{figure}[ht]
\vspace{-1.75cm}
\includegraphics[width=8.5cm]{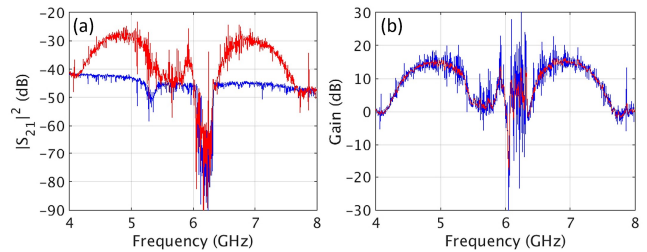}
\vspace{-1.75cm}
\caption{\small Gain-bandwidth measurements from two cascaded ``leaf" amplifiers with embedded resonators. (a) Transmission $|S_{21}|^{2}$ with the pump off (blue) and the pump on (red). (b) Gain profile. The measured gain curve is shown in blue, while a smoothed gain curve is shown in red.}
\label{fig:leaf_gain}
\end{figure}

The measured performance of the cascaded ``leaf" amplifiers using resonator phase shifters is shown in Fig. \ref{fig:leaf_gain}. The transmission through the device with the pump off (blue) and the pump on (red) is illustrated in Fig. \ref{fig:leaf_gain}a. As indicated by the pump-off transmission, the resonant frequency of the embedded resonators falls at approximately 6.2 GHz. By injecting a pump tone at 5.92 GHz with 60 $\mu$W of power, we achieve a peak gain of 15 dB, with a 3 dB double-sided bandwidth of 1 GHz as shown in Fig. \ref{fig:leaf_gain}c. The bandwidth is smaller than the periodically-loaded amplifier. While the cause is under investigation, we speculate it may be related to two factors. First, the stopband near the pump is wider (300 MHz for the ``leaf" amplifier vs. 160 MHz for the ``fishbone" amplifier) which consumes wider bandwidth. Second, because the cutoff frequency is lower (55 GHz vs. 318 GHz), the intrinsic low-pass dispersion introduces greater phase mismatch, which reduces the gain at a frequency detuned from the pump. We also notice a dip at 5.3 GHz in the pump-off transmission of Fig. \ref{fig:leaf_gain}a. The unexpected dip is likely due to a parasitic mode and we believe it is not intrinsic to the design. The gain ripple at the top of the gain curve, about $\pm2$ dB, is similar to the ``fishbone" design.




It is useful to compare the artificial transmission line amplifier design with the conventional CPW design \cite{pappas_2013}. The key aspect is the large interdigitated capacitors introduced in the artificial transmission line architectures, which has effectively increased the capacitance per unit length by a factor of 16 as compared to the distributed capacitance in the CPW counterpart. This, in turn, has reduced the characteristic impedance (to 50 ohms) and phase velocity by a factor of 4. As the pump power $P_{p}=I_{rms}^{2}Z_{0}$ varies linearly in the characteristic impedance, the reduced characteristic impedance results in a factor of 4 lower pump power for the same inductor width and nonlinearity. The reduced phase velocity means a shorter wavelength and a shorter physical length. Even though the ``fishbone" and ``leaf" amplfiers have smaller phase length than the CPW (140 wavelengths vs 400 for the CPW), the gain is approximately the same. This suggests that the amplifier is pumped to a higher nonlinearity (the time average of $I^{2}/I_{*}^{2}$ is increased to a larger value), and the gain media is utilized more efficiently, explaining why the reduction in pump power is not a factor of 4, but 2-3. The artificial transmission line architecture also improves the fabrication yield because the 10$\times$ reduction in length makes it less likely for the amplifier to have defects that break the center inductor strip. Also, defects that break a few IDC fingers will not prevent the amplifier from operating. 

In conclusion, we have demonstrated traveling-wave parametric amplifiers based on the nonlinear kinetic inductance of NbTiN thin films and an artificial lumped-element transmission line architecture. These amplifiers, using two different phase-matching schemes, exhibit 15 dB gain over a few GHz of bandwidth centered at a pump frequency near 6 GHz. Compared to the previous CPW design, our amplifiers provide similar high gain and large bandwidth, but with an order of magnitude reduction in trace length and a factor of 2-3 reduction in pump power. Their intrinsic 50 ohm characteristic impedance has greatly improved the impedance matching and reduced the amplitude of the gain ripple. Future investigations will include measurements of the amplifier noise. We expect that the reduced pump power will help to mitigate thermal noise which is known to degrade the performance in previous generations of NbTiN parametric amplifiers\cite{day_2014}. The amplifier architecture can easily accommodate narrower inductor strips or even nanowires, which is impractical for a CPW design. We may also construct the amplifier using a different superconducting nitride, such as TiN, which possesses a lower critical current density and may therefore provide a route for reduced pump power. Additionally, improved understanding of the kinetic inductance nonlinearity may lead to further device optimization\cite{semenov}. The amplifier presented here may be used in important low-temperature applications such as the readout of thousands of detectors or qubits.

\vspace{1 em}

The amplifier devices were fabricated in the NIST-Boulder microfabrication facility. This work was supported by NASA Astrophysics Research and Analysis Program Grant \#NNH12ZDA001N. S. Chaudhuri was supported by a NASA Space Technology Research Fellowship, NASA Grant \#NNX14AM48H. 


\newpage

\end{document}